\documentclass[a4paper,11pt]{article}
\usepackage{amsmath}
\usepackage{amssymb}
\usepackage[pdftex]{graphicx}
\usepackage{epstopdf}
\usepackage{longtable}
\usepackage{subfigure}
\usepackage{natbib}
\usepackage[pdftex, pdftitle={Recombinations of Busy Beaver Machines},
pdfauthor={Norbert Batfai},
pdfkeywords={Busy Beaver, Placid Platypus},
pdfstartview=FitB]{hyperref}
\usepackage{amssymb,amsmath,amsthm,amsfonts}
\usepackage{tikz}

\newtheorem{definition}{Definition}

\newtheorem{remark}{Remark}
\newtheorem{example}{Example}

\newcommand{\keywords}[1]{\par\addvspace\baselineskip\noindent\textbf{Keywords: }\textit{#1}.}

\author{Norbert B\'atfai
\\University of Debrecen
\\Department of Information Technology
\\\texttt{batfai.norbert@inf.unideb.hu}}
\title{Recombinations of Busy Beaver Machines}
\begin{document}
\maketitle

\begin{abstract}
Many programmers belive that Turing-based machines cannot think. We also believe in this, however it is interesting to note that the most sophisticated machines are not programmed by human beings. We have only discovered them. In this paper, using well-known Busy Beaver and Placid Platypus machines, we generate further very similar, but not exactly the same machines. We have found a recombinated $BB_5$ machine which can make $70.740.809$ steps before halting.
\keywords{Busy Beaver, Placid Platypus}
\end{abstract}

\tableofcontents

\section{Introduction}

Many people and many programmers belive that a computer program cannot think. We also believe in this, because Turing machines have no consciousness, they cannot flip a coin, and so on. But unfortunately, nowadays we are not able to be more precise in this question. However it is interesting to note that the most sophisticated machines are not programmed by us. We have only discovered them.

In the following  we have restricted ourselves to the case of Busy Beaver machines because radically advanced machines may be found among these Turing machines. We employ the notations introduced in \citep{batfai-2009}. The Busy Beaver problem, introduced by Tibor Rad\'o more than 40 years ago, is to find the $n$-state, binary tape Turing machine that is starting with the empty word and has written the greatest number of ones on the tape. The Placid Platypus problem, introduced by Harland in \citep{PlacidPlatypus}, is the inverse problem of Busy Beaver. It is to find the fewest number of states binary tape Turing machine that is starting with the empty word and has written $n$ ones on the tape.

For example the $5$-state winner candidate (Marxen-Buntrock) machine is a very special machine. It can make $47.176.869$\footnote{The variant of definition of Turing machine used by us may changes the Busy Beaver functions. For example, Rad\'o's $S(n)$ may equals to $S(n)$ or $S(n)-1$. For the details we refer to \citep[pp. 5]{batfai-2009}.} steps before halting. Let us try it. This machine cannot be created by any usual programming methods. 
It works in such a way that we would never be able to develop it from entirely scratch.

In our paper, we simply combine some well-known machines to produce new machines. This work was suggested by a paper from Michel. He wrote in \citep[pp. 11]{michel-2009} that according to Marxen in range from $4096$ to $4098$ there are no further machines. 

\section{Recombinations of machines}

\subsection{Combining well-known machines}

\begin{definition}[Recombinations]
Let $T_{i_1}, \dots T_{i_k}$ be $BB_n$ Turing machines.
A recombination of machines $T_{i_1}, \dots T_{i_k}$
is a $(2k-1)$-tuple $T_{i_{k+1}}=(T_{i_1}, \dots T_{i_k},$ $v_{j_1}, \dots , v_{j_{k-1}})$ 
where  $T_{i_1}, \dots T_{i_k}$  are the source machines, $0 \le  v_{j_{m}} \le  v_{j_{m+1}} \le 2n-1$ and the partial transition function of recombined machine $T_{i_{k+1}}$ is defined as 
\[
\begin{aligned}
(
 &f^0_{T_{i_1}} \rightarrow t^0_{T_{i_1}}, f^1_{T_{i_1}} \rightarrow t^1_{T_{i_1}}, 
\dots, f^{v_{j_1}-1}_{T_{i_1}} \rightarrow t^{v_{j_1}-1}_{T_{i_1}}\\
 &f^{v_{j_1}}_{T_{i_2}} \rightarrow t^{v_{j_1}}_{T_{i_2}}, f^{v_{j_1}+1}_{T_{i_2}} \rightarrow t^{v_{j_1}+1}_{T_{i_2}}, 
\dots, f^{v_{j_2}-1}_{T_{i_2}} \rightarrow t^{v_{j_2}-1}_{T_{i_2}}\\
 & \vdots\\
 &f^{v_{j_{k-1}}}_{T_{i_k}} \rightarrow t^{v_{j_{k-1}}}_{T_{i_k}}, f^{v_{j_{k-1}+1}}_{T_{i_k}} \rightarrow t^{v_{j_{k-1}+1}}_{T_{i_k}},
\dots, f^{2n-1}_{T_{i_k}} \rightarrow t^{2n-1}_{T_{i_k}} 
)
\end{aligned}
\]
where  $f^i_T \rightarrow t^i_T$ is the i-th transition rule of machine T. 
\end{definition}

\begin{example}
Let $T_i$, $T_j$ and $T_k$ be three $BB_5$ Turing machines.
A recombination of machines $T_i$, $T_j$ and $T_k$
is a quintuple $R=(T_i, T_j, T_k, u, v)$ 
where  $T_i$, $T_j$ and $T_k$ are the source machines, $0 \le u \le v \le 9$. 
This machine $R$ is shown in Figure \ref{recomb}.

\begin{figure}[htp]
\begin{center}
  \begin{tikzpicture}
    \draw (0,0) rectangle (5,1)
    rectangle (0,2)
    rectangle (5,3);
    \node at (2.5,0.5) {rules from the k-th machine};
    \node at (2.5,1.5) {rules from the j-th machine};
    \node at (2.5,2.5) {rules from the i-th machine};
    \draw (0,1) -- (5.5,1);
    \draw (0,2) -- (5.5,2);
    \node at (6,2) {u};
    \node at (6,1) {v};
  \end{tikzpicture}
\end{center}
\caption{Rules of the recombinated machine $R$.}
\label{recomb}
\end{figure}
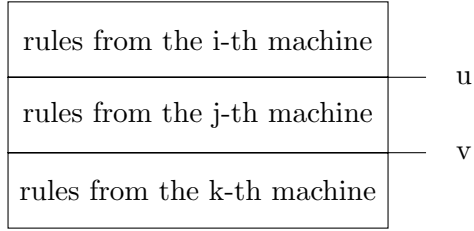
\end{example}
\subsection{The starting machines}

\begin{figure}[htp]
\centering
\subfigure[4097, 47.176.869, (9, 0, 11, 1, 15, 2, 17, 3, 11, 4, 23, 5, 24, 6, 3, 7, 21, 9, 0)]{\label{mb1}\includegraphics[scale=0.38]{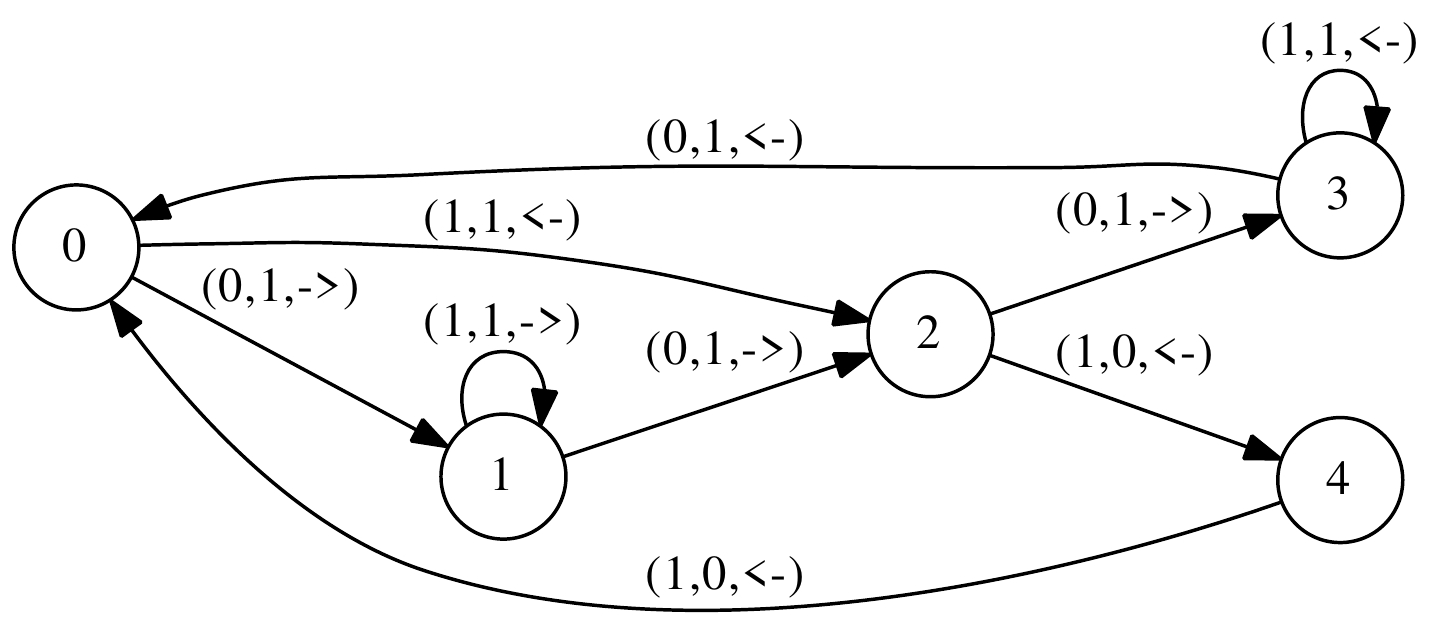}} %
\subfigure[4096, 23.554.763, (9, 0, 11, 1, 18, 2, 15, 3, 23, 4, 3, 5, 15, 7, 29, 8, 5, 9, 8)]{\label{mb2}\includegraphics[scale=0.38]{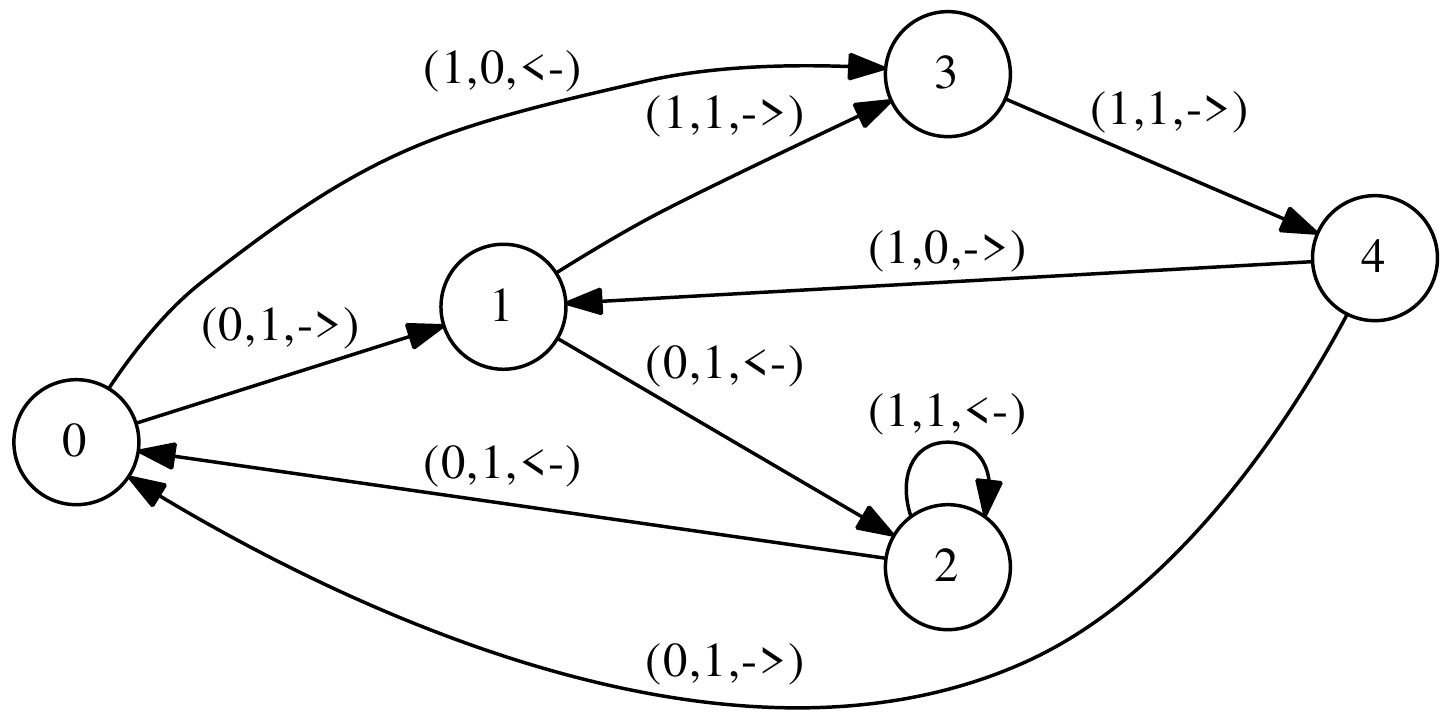}} \\ %
\subfigure[4095, 11.804.909, (9, 0, 11, 1, 5, 2, 15, 3, 20, 4, 3, 5, 15, 7, 29, 8, 24, 9, 11)]{\label{mb3}\includegraphics[scale=0.33]{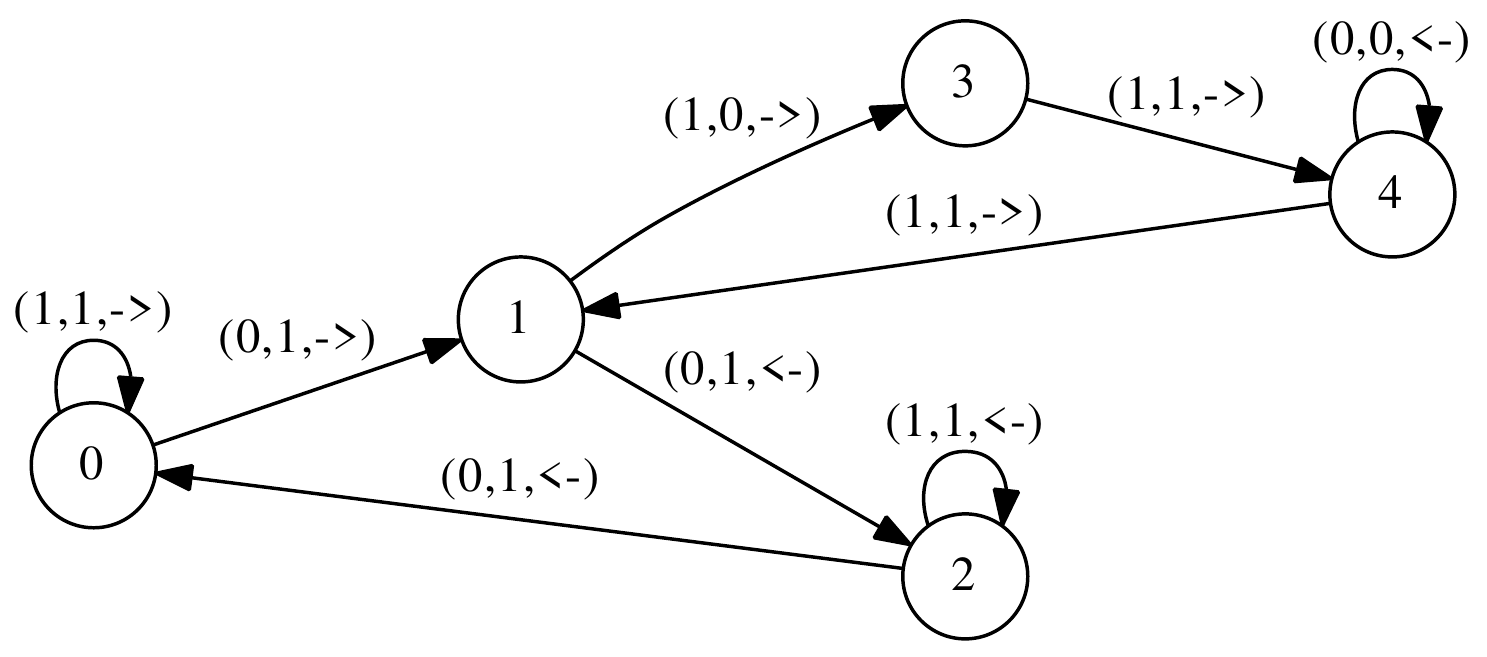}} %
\subfigure[4095, 11.804.895, (9, 0, 11, 1, 5, 2, 15, 3, 20, 4, 3, 5, 15, 7, 29, 8, 15, 9, 11)]{\label{mb4}\includegraphics[scale=0.33]{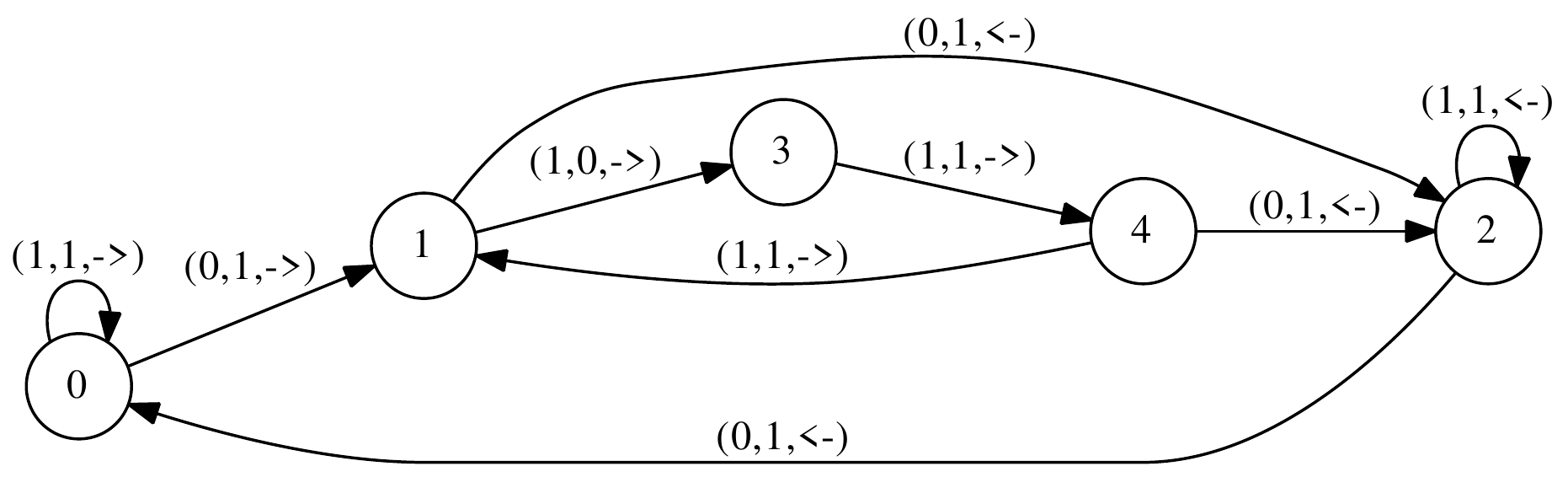}} \\ %
\subfigure[4097, 11.798.825, (9, 0, 11, 1, 5, 2, 15, 3, 9, 4, 5, 5, 21, 6, 5, 7, 27, 9, 12)]{\label{mb5}\includegraphics[scale=0.33]{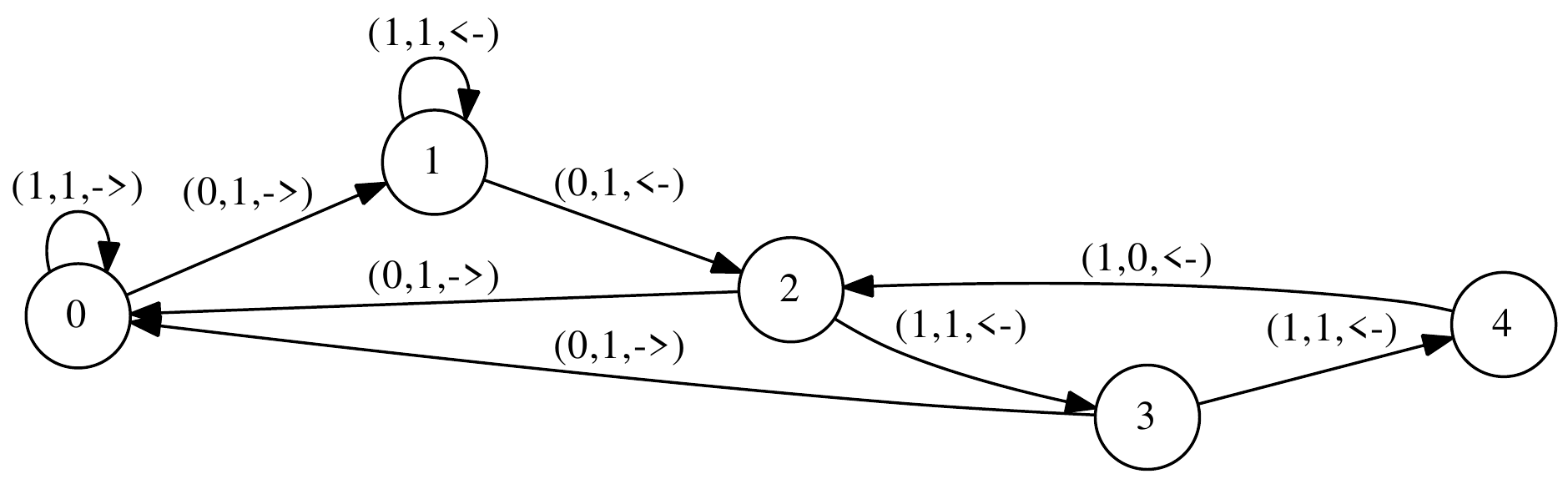}} %
\subfigure[4096, 11.798.795, (9, 0, 11, 1, 5, 2, 15, 3, 23, 4, 3, 5, 15, 7, 26, 8, 15, 9, 11)]{\label{mb6}\includegraphics[scale=0.33]{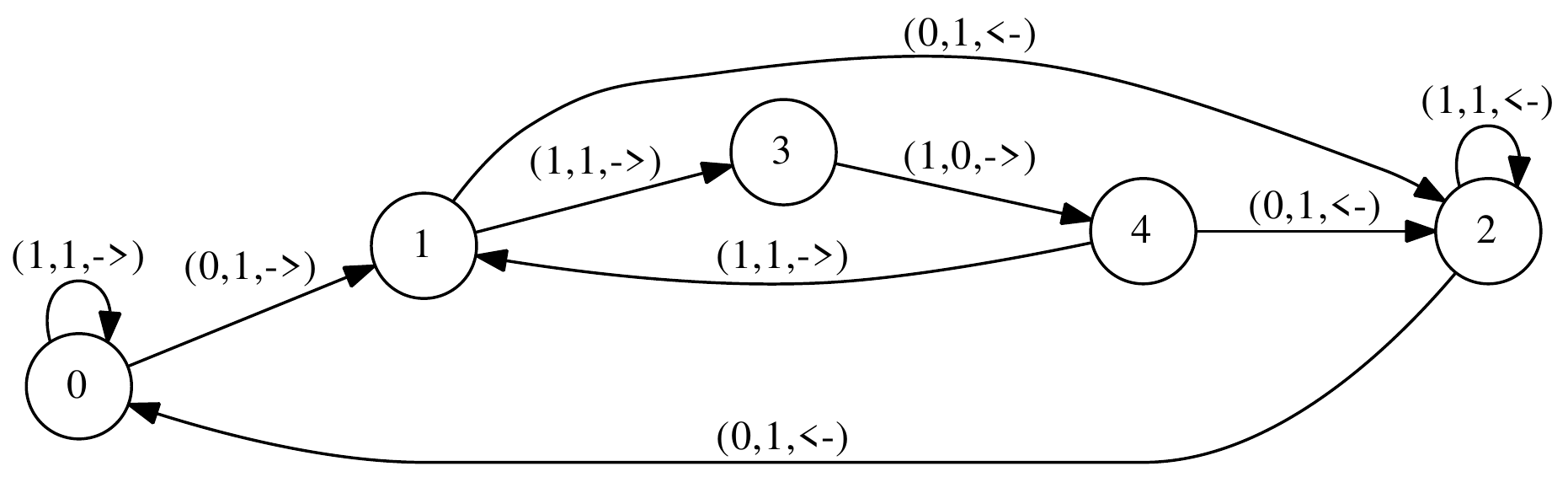}} %
\caption{Marxen and Buntrock's winner candidate machines, \# of 1's, \# of steps, name of the machine.}
\label{mb}
\end{figure}

\begin{figure}[htp]
\centering
\subfigure[1471, 2.358.063, (9, 0, 11, 2,  15, 3, 17, 4,  26, 5,  18, 6,  15, 7,  6,  8, 23, 9,  5 )]{\label{us1}\includegraphics[scale=0.38]{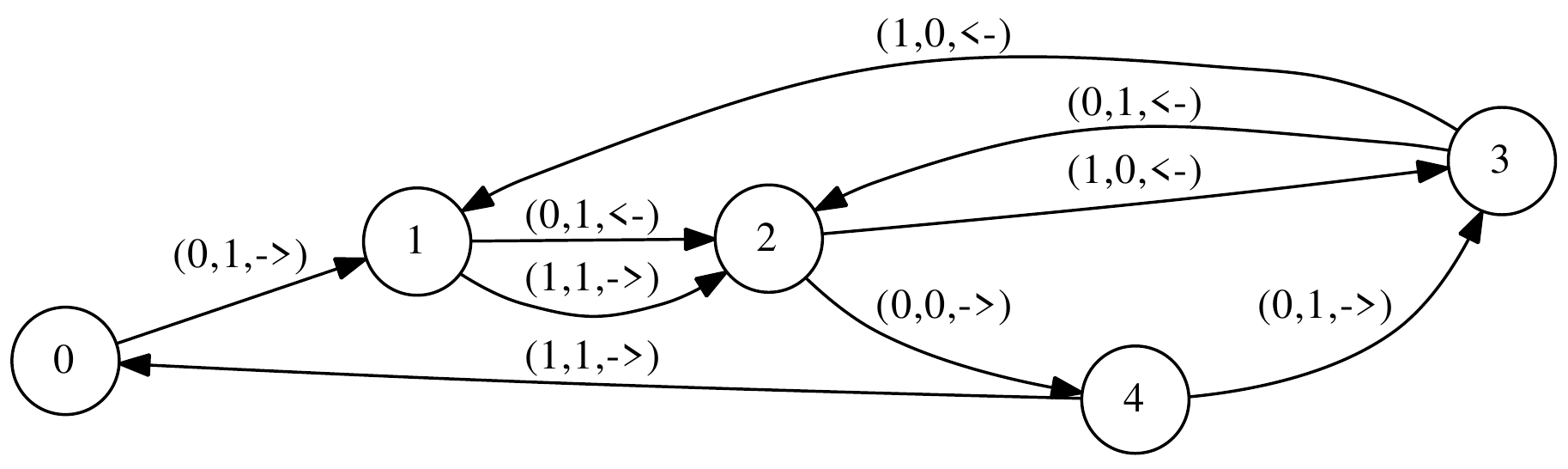}} %
\subfigure[1915, 2.133.491, (9, 0, 11, 1,  15, 2, 0,  3,  18, 4,  3,  6,  9,  7,  29, 8, 20, 9,  8 )]{\label{us2}\includegraphics[scale=0.38]{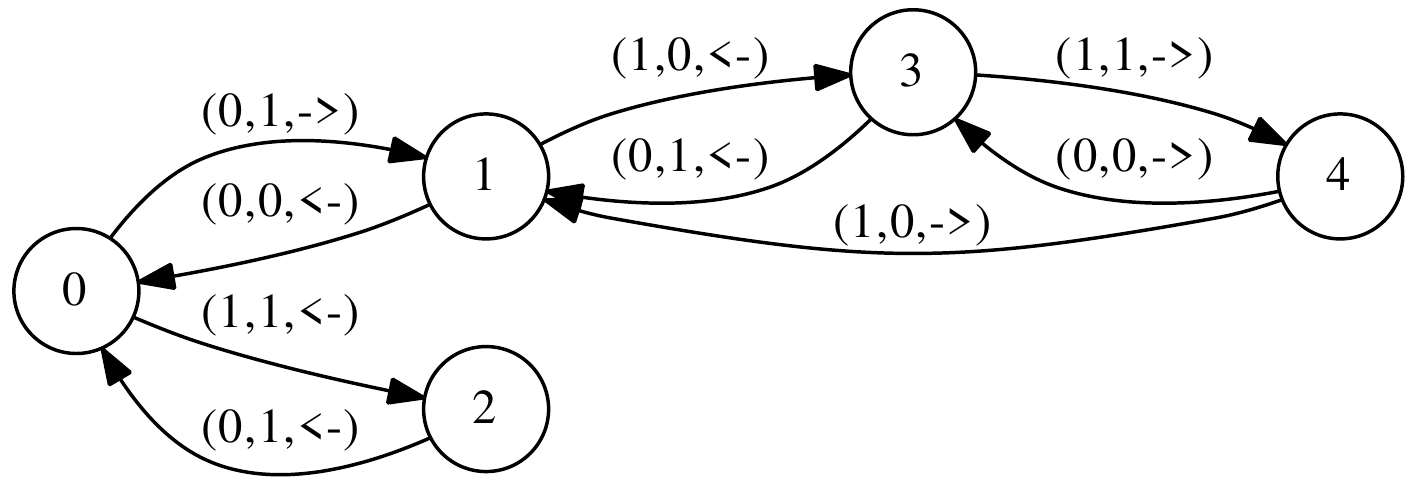}} \\ %
\subfigure[ 501, 134.466, (9, 0, 11, 1,  12, 2, 17, 3,  23, 4,  3,  5,  8,  6,  26, 8, 15, 9,  5)]{\label{us3}\includegraphics[scale=0.33]{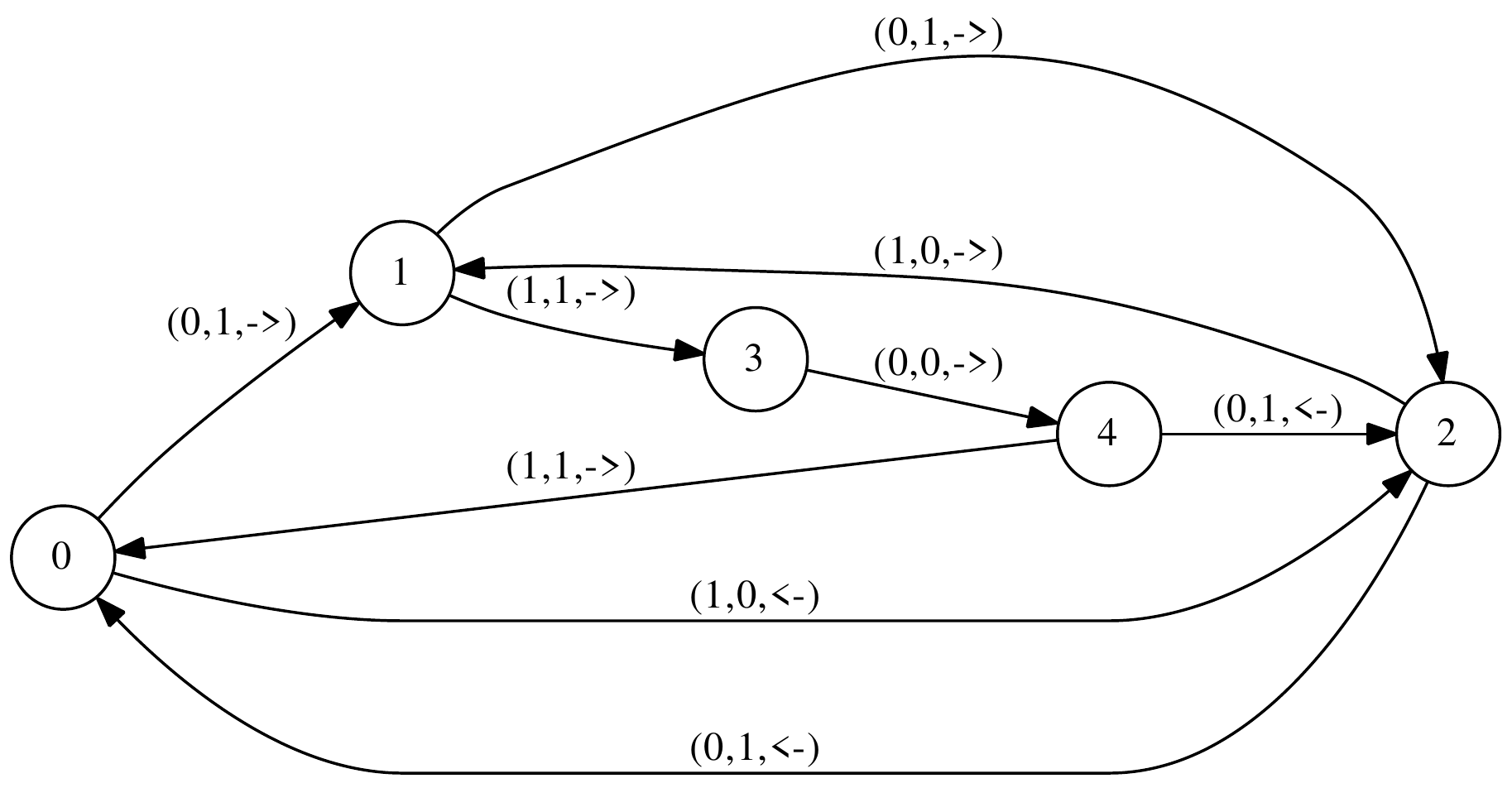}} %
\caption{Uhing, and Schult machines, \# of 1's, \# of steps, name of the machine.}
\label{us}
\end{figure}

We have started some well known $BB_5$ machines. The first six, the two winner candidate $\mathcal{M}_{PP}(4097)$\footnote{The variant of definition of Turing machine used by us may changes the Busy Beaver functions. For example, Rad\'o's $\Sigma(n)$ may equals to $\Sigma(n)$ or $\Sigma(n)-1$. For the details we refer to \citep[pp. 5]{batfai-2009}.}, two $\mathcal{M}_{PP}(4096)$ and two $\mathcal{M}_{PP}(4095)$ machines shown in Figure \ref{mb} are discovered by Marxen and Buntrock \citep{michel-2009}. The next three, $\mathcal{M}_{PP}(1471)$, $\mathcal{M}_{PP}(1915)$ and $\mathcal{M}_{PP}(501)$ machines  shown in Figure \ref{us} are found by Uhing and Schult \citep{michel-2009}. Finally, the residual some $\mathcal{M}_{PP}(160)$, $\mathcal{M}_{PP}(32)$, $\mathcal{M}_{PP}(26)$, $\mathcal{M}_{PP}(21)$ and from $\mathcal{M}_{PP}(19)$ to $\mathcal{M}_{PP}(2)$ machines are found by our own programs \citep{batfai-2009}.

\clearpage
\subsection{Recombinations of machines}

In this section, from Figure \ref{own}, we show machines which are recombinated by our program from the starting machines. These are very similar to the starting machines, but different ones. 

\begin{figure}[htp]
\centering                      
\subfigure[4096, 11.792.681, (9, 0, 11, 1, 5, 2, 15, 3, 23, 4, 3, 5, 15, 7, 29, 8, 5, 9, 8)]{\label{recomb4096}\includegraphics[scale=0.38]{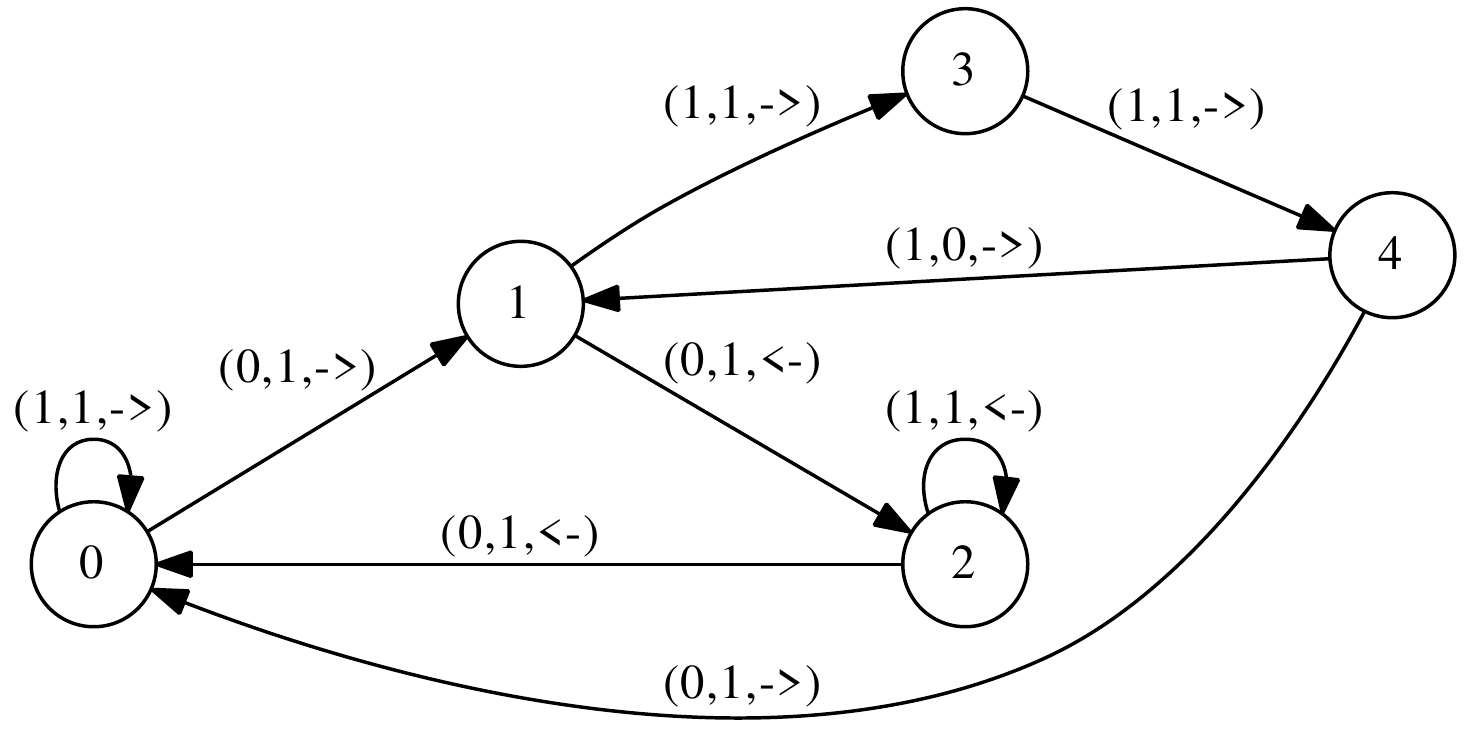}} 
\subfigure[4096, 11.792.723, (9, 0, 11, 1, 5, 2, 15, 3, 23, 4, 3, 5, 15, 7, 29, 8, 24, 9, 8)]{\label{recomb4096_2}\includegraphics[scale=0.38]{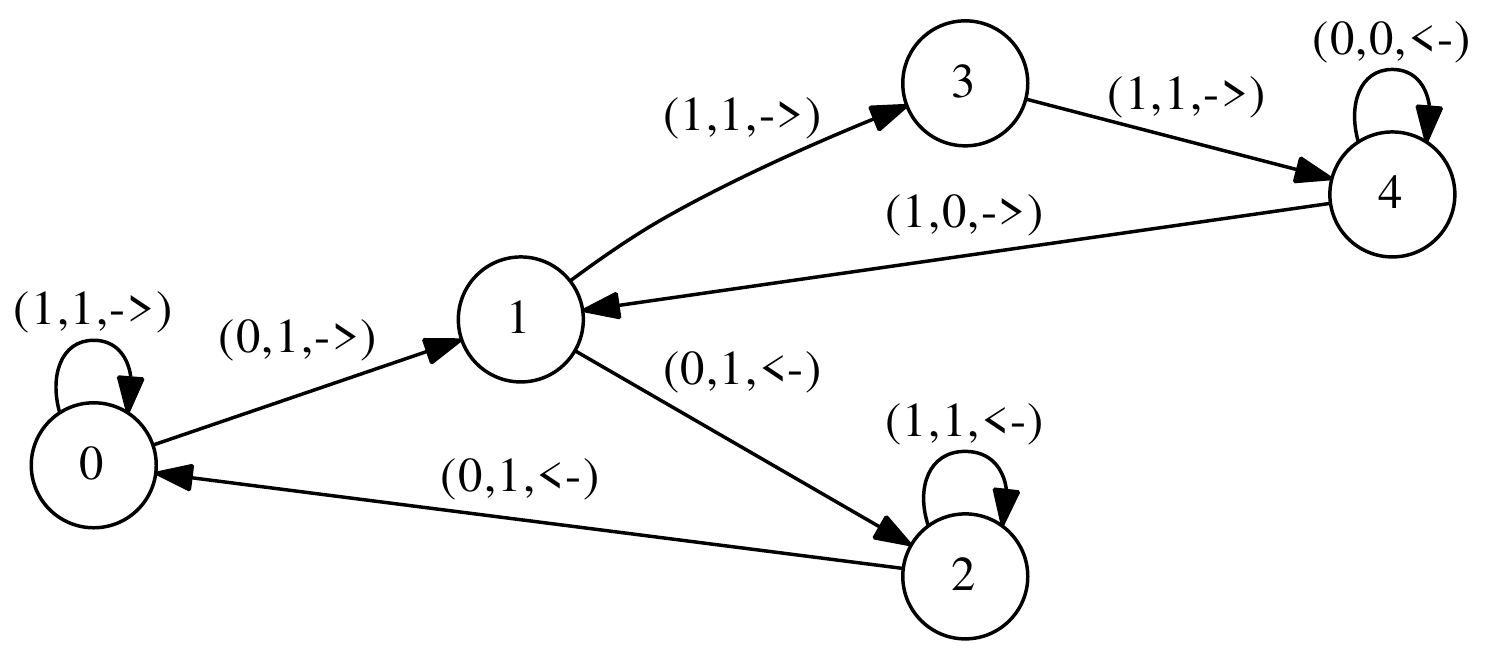}}\\ 
\subfigure[4096, 11.803.885, (9, 0, 11, 1, 5, 2, 15, 3, 23, 4, 3, 5, 15, 7, 26, 8, 15, 9, 1)]{\label{recomb4096_3}\includegraphics[scale=0.38]{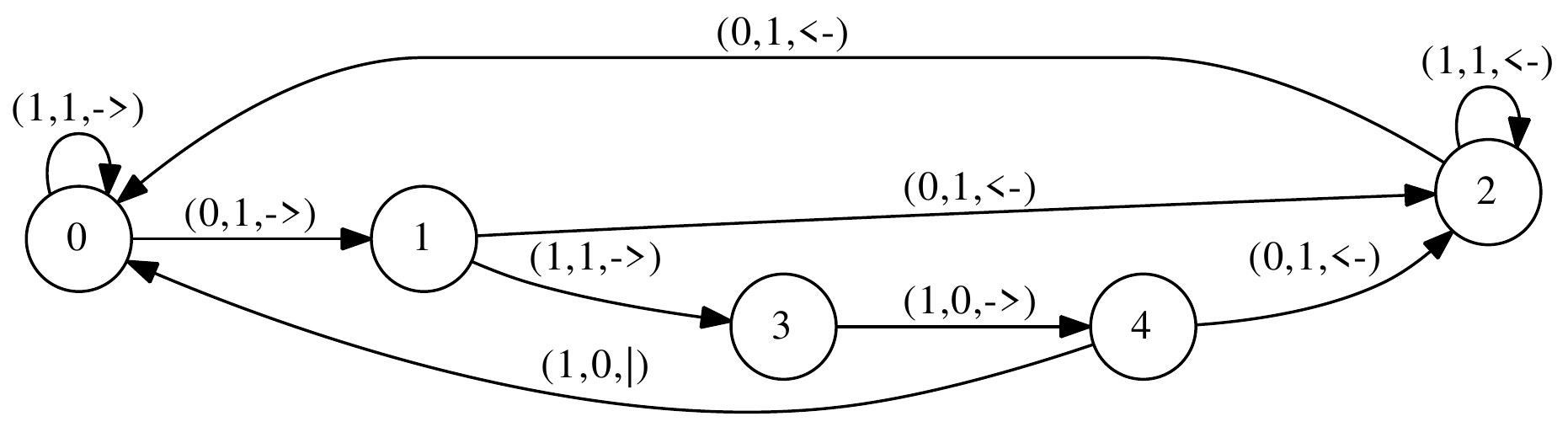}} 

\caption{Recombinations of $BB_5$ machines, $\mathcal{M}_{PP}(4096)$ machines, \# of 1's, \# of steps, name of the machine.}
\end{figure}

\begin{example}
The machine in figure \ref{recomb4096_2}, is found on the following ''evolutionary tree'' shown in Figure \ref{tree}. The names of the contained Turing machines may be decrypted using row indexes in code snipets in Appendix \ref{codesnips}.

\tikzstyle{level 1}=[level distance=1.5cm, sibling distance=8cm]
\tikzstyle{level 2}=[level distance=1.5cm, sibling distance=6cm]
\tikzstyle{level 3}=[level distance=1.5cm, sibling distance=4cm]
\tikzstyle{level 4}=[level distance=1.5cm, sibling distance=2cm]

\begin{figure}[htp]
  \begin{center}
    \begin{tikzpicture}[grow=up] 
      \node {(16,32,16,6,8)}
      child {
        node {(5,1,6)}        
        edge from parent 
        node[above] {16}
      }
      child {
        node {(1,25,6,4,9)}        
        child {
          node {(6,18,20,3,4)}        
          child {
            node {(5,15,5,7,8)}        
            child {
              node {(3,9,5,7,8)}        
              edge from parent 
              node[right] {15}
            }
            edge from parent 
            node[above] {18}
          }
          child {
            node {(14,0,2,4,5)}        
            child {
              node {(4,1,12,2,5)}        
              edge from parent 
              node[right] {14}
            }
            edge from parent 
            node[above] {20}
          }          
          edge from parent 
          node[right] {25}
        }
        edge from parent 
        node[above] {32}
      };
    \end{tikzpicture}
  \end{center}
  \caption{Evolutionary tree of machine (9, 0, 11, 1, 5, 2, 15, 3, 23, 4, 3, 5, 15, 7, 29, 8, 24, 9, 8) shown in Figure \ref{recomb4096_2}.}
\label{tree}
\end{figure}
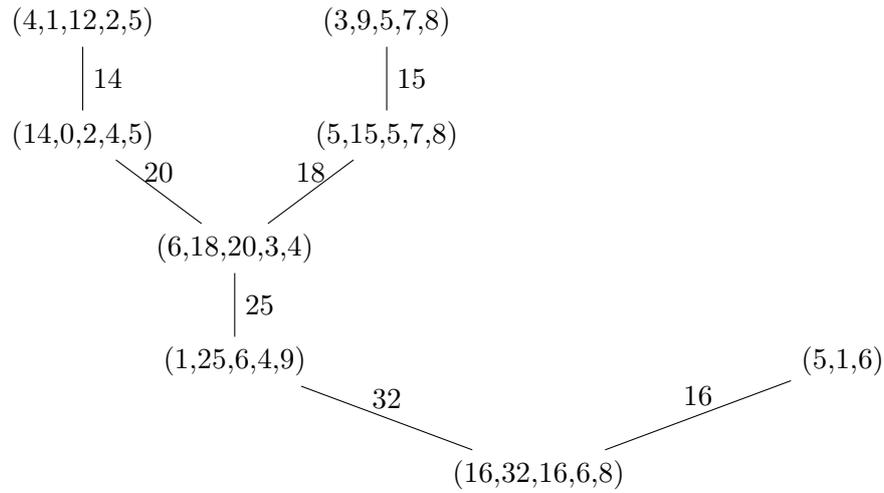
\end{example}

\clearpage
\subsubsection{$\mathcal{M}_{PP}(4097)$}

\begin{figure}[htp]
\centering                      
\subfigure[4097, 11.801.882, (9, 0, 11, 1, 5, 2, 15, 3, 9, 4, 19, 5, 21, 6, 5, 7, 27, 9, 12)]{\label{recomb4097}\includegraphics[scale=0.33]{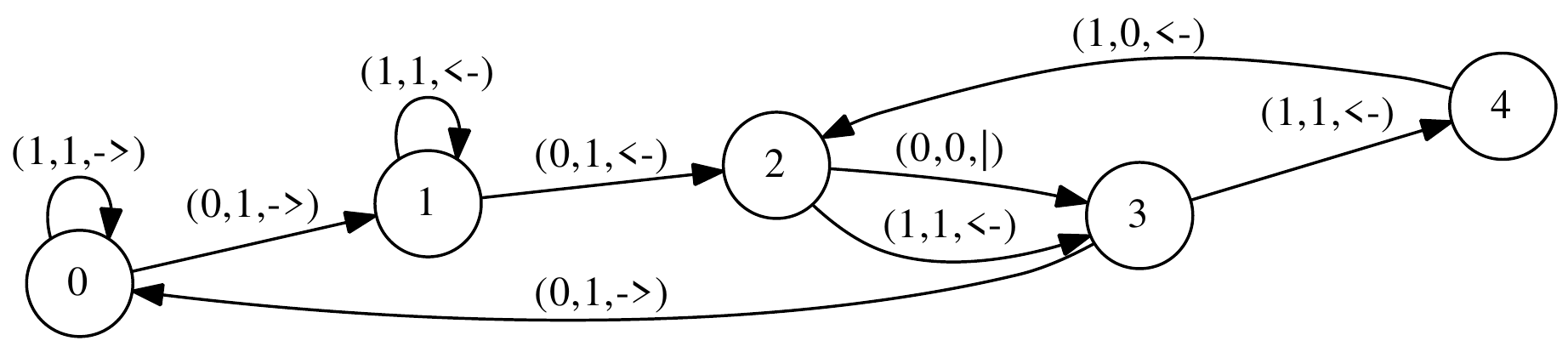}} 
\subfigure[4097, 11.798.832, (9, 0, 11, 1, 5, 2, 15, 3, 9, 4, 5, 5, 21, 6, 4, 7, 27, 9, 12)]{\label{recomb4097_2}\includegraphics[scale=0.33]{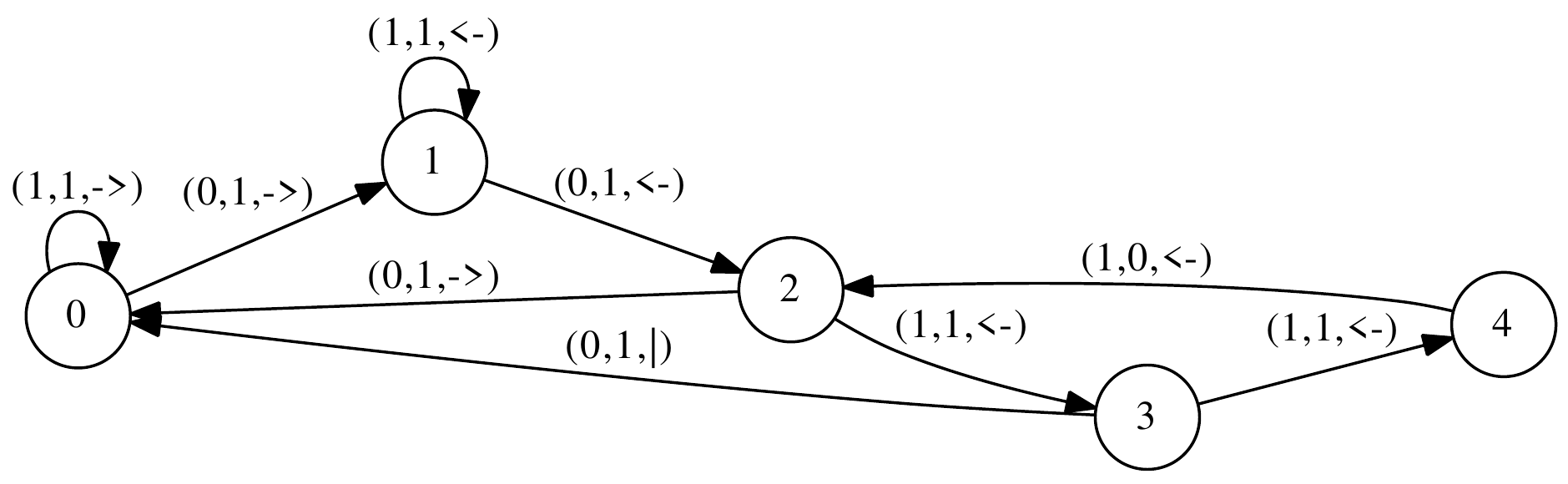}} \\ 
\subfigure[4097, 70.740.809, (9, 0, 11, 1, 15, 2, 17, 3, 1, 4, 23, 5, 24, 6, 3, 7, 21, 9, 0)]{\label{recomb4097_3}\includegraphics[scale=0.35]{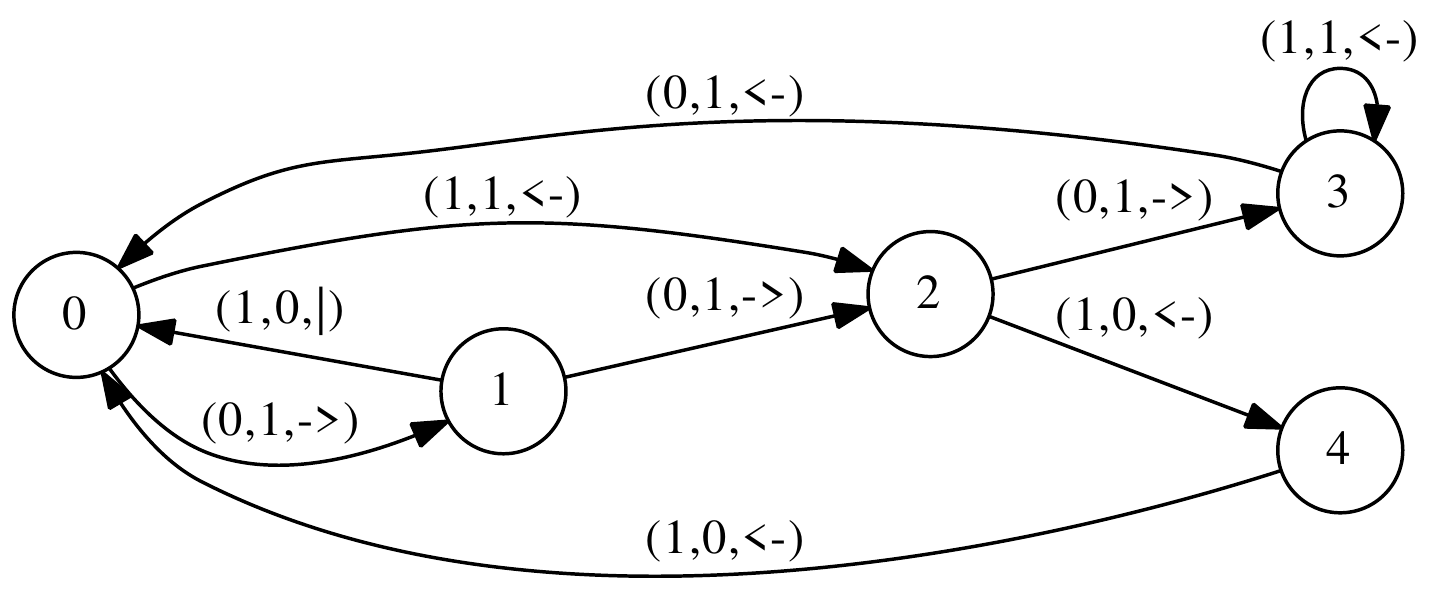}} 
\subfigure[4097, 17.689.051, (9, 0, 11, 1, 13, 2, 15, 3, 9, 4, 5, 5, 21, 6, 5, 7, 27, 9, 12)]{\label{recomb4097_4}\includegraphics[scale=0.35]{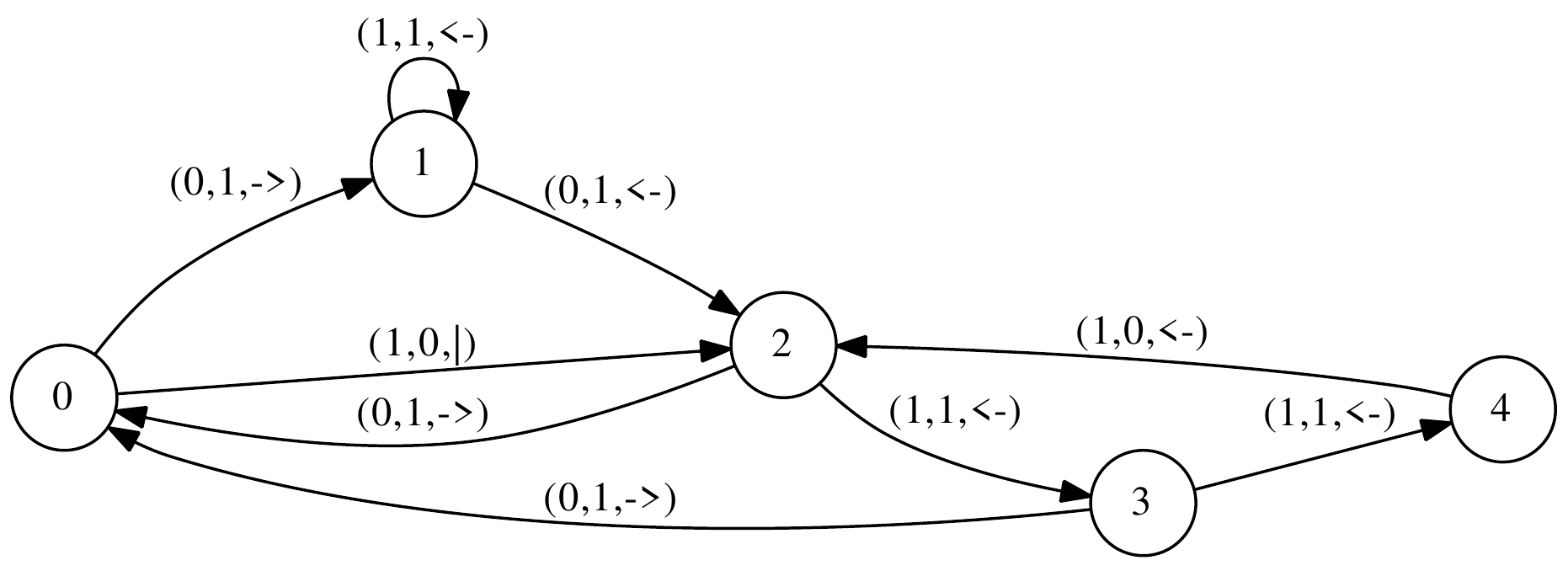}} \\
\subfigure[4097, 23.582.334, (9, 0, 11, 1, 13, 2, 15, 3, 9, 4, 19, 5, 21, 6, 5, 7, 27, 9, 12)]{\label{recomb4097_5}\includegraphics[scale=0.33]{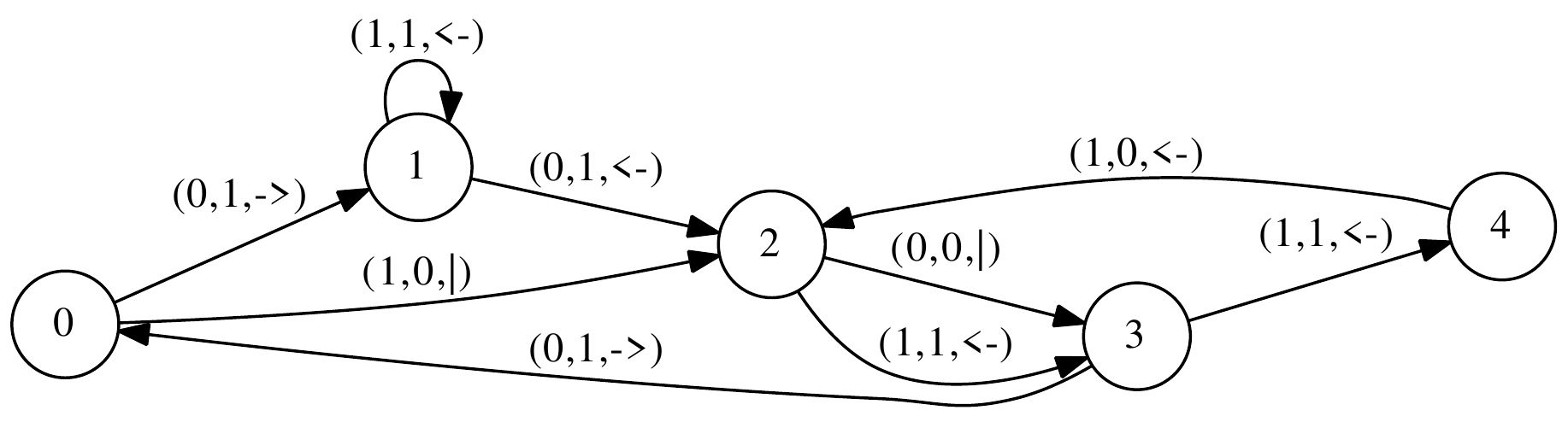}} 
\subfigure[4097, 17.689.065, (9, 0, 11, 1, 13, 2, 15, 3, 9, 4, 5, 5, 21, 6, 4, 7, 27, 9, 12)]{\label{recomb4097_6}\includegraphics[scale=0.33]{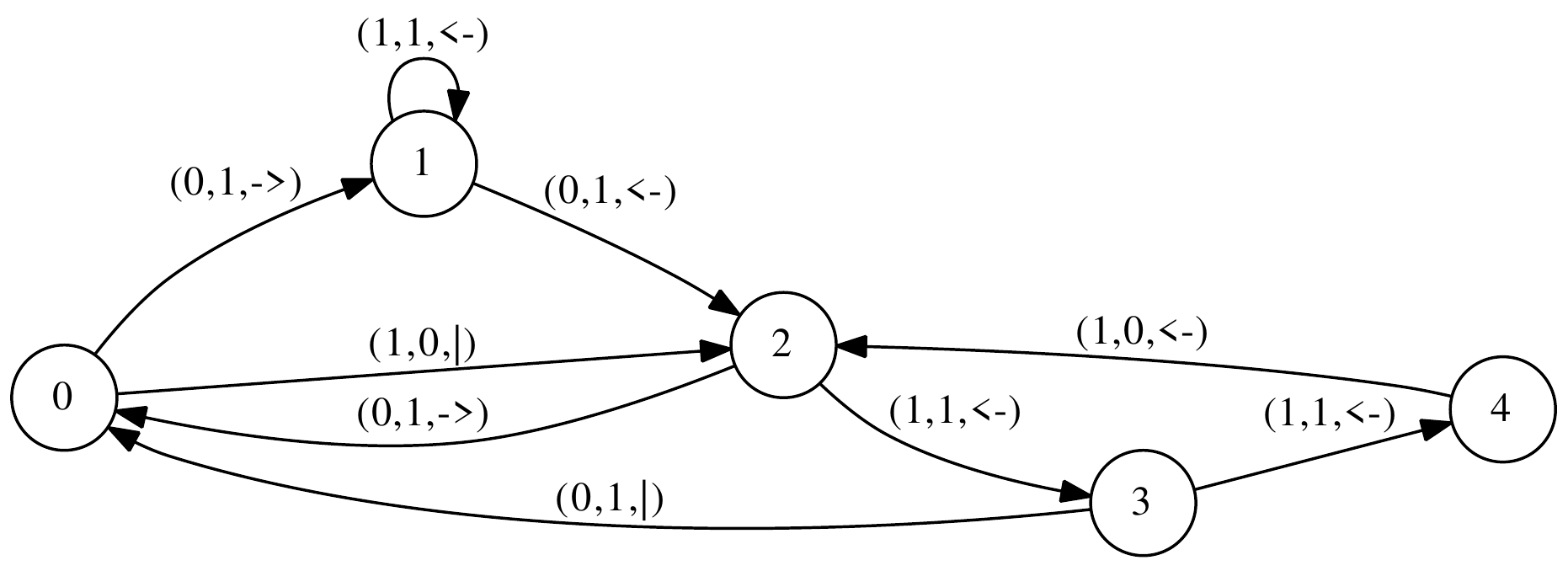}} 
\subfigure[4097, 11.804.946, (9, 0, 11, 1, 5, 2, 15, 3, 9, 4, 19, 5, 21, 6, 4, 7, 27, 9, 12)]{\label{recomb4097_7}\includegraphics[scale=0.33]{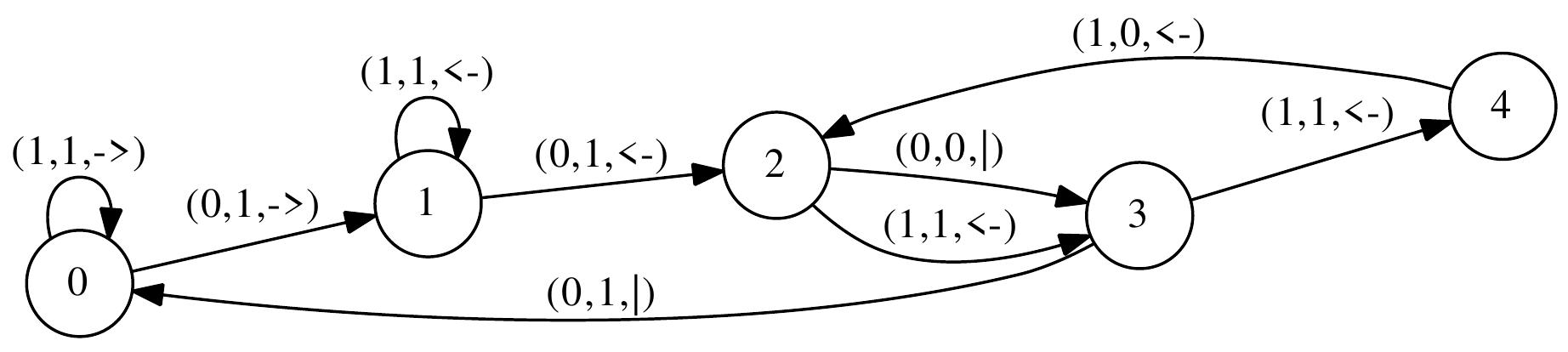}} \\
\caption{Recombinations of $BB_5$ machines, $\mathcal{M}_{PP}(4097)$ machines, \# of 1's, \# of steps, name of the machine.}
\label{own}
\end{figure}

\begin{remark}
\label{corollary}
We should remark that we differ from Rad\'o's original Turing machine model. We have used an unusual Turing machine model for Busy Beaver problem. On the one hand, the definition of Turing machine used by us may changes the Busy Beaver function, because we do not use an extra halting state, for the details we refer to \citep[pp. 5]{batfai-2009}. And on the other hand, staying on the same tape cell on a given step is permitted by our model.    

If we compare Marxen and Buntrock's original machine shown in Figure \ref{mb1} and the recombined machine shown in Figure \ref{recomb4097_3} we will notice that in case of these machines the transition rule $(1,1) \rightarrow (1,1,\rightarrow)$ may be replaced with the following two "unusual" rules $(0,0) \rightarrow (1,1,\rightarrow)$ and $(1,1) \rightarrow (0,0,|)$.

\begin{figure}[htp]
\centering 
\begin{minipage}[b]{4 cm}
\subfigure[$(1,1) \rightarrow (1,1,\rightarrow)$]{\label{repl1}\includegraphics[scale=.5]{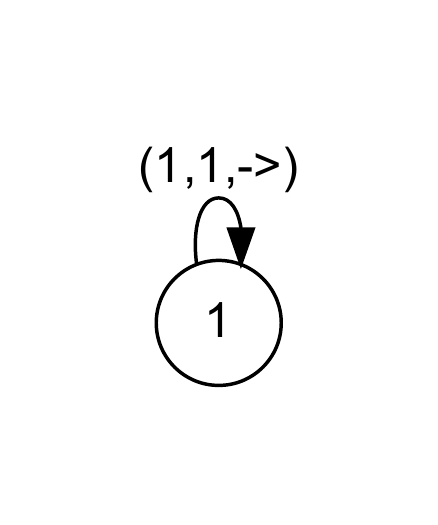}}
\end{minipage}
\begin{minipage}[b]{4 cm}
\subfigure[$(0,0) \rightarrow (1,1,\rightarrow)$ and $(1,1) \rightarrow (0,0,|)$]{\label{repl2}\includegraphics[scale=.5]{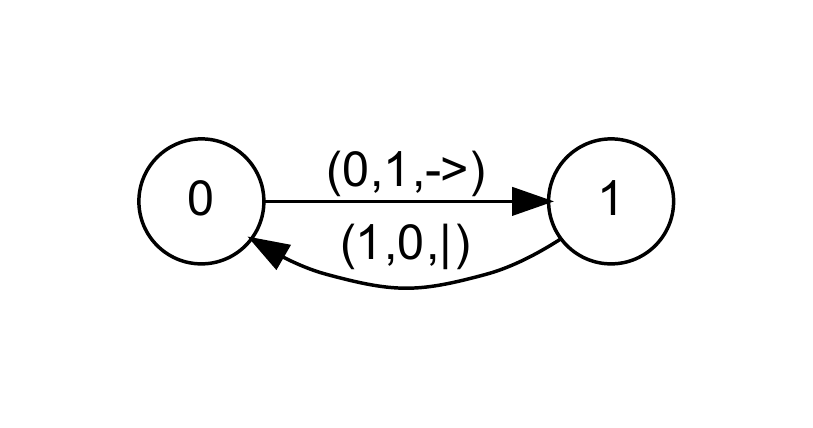}} 
\end{minipage}                     
\caption{Stepping over a block of ones.}
\label{repl}
\end{figure}

\end{remark}

\clearpage
\subsubsection{$\mathcal{M}_{PP}(4096)$}

\clearpage
\subsubsection{$\mathcal{M}_{PP}(4095)$}

\begin{figure}[htp]
\centering                      
\subfigure[4095, 11.815.075, (9, 0, 11, 1, 5, 2, 15, 3, 20, 4, 3, 5, 15, 7, 29, 8, 15, 9, 0)]{\label{recomb4095}\includegraphics[scale=0.33]{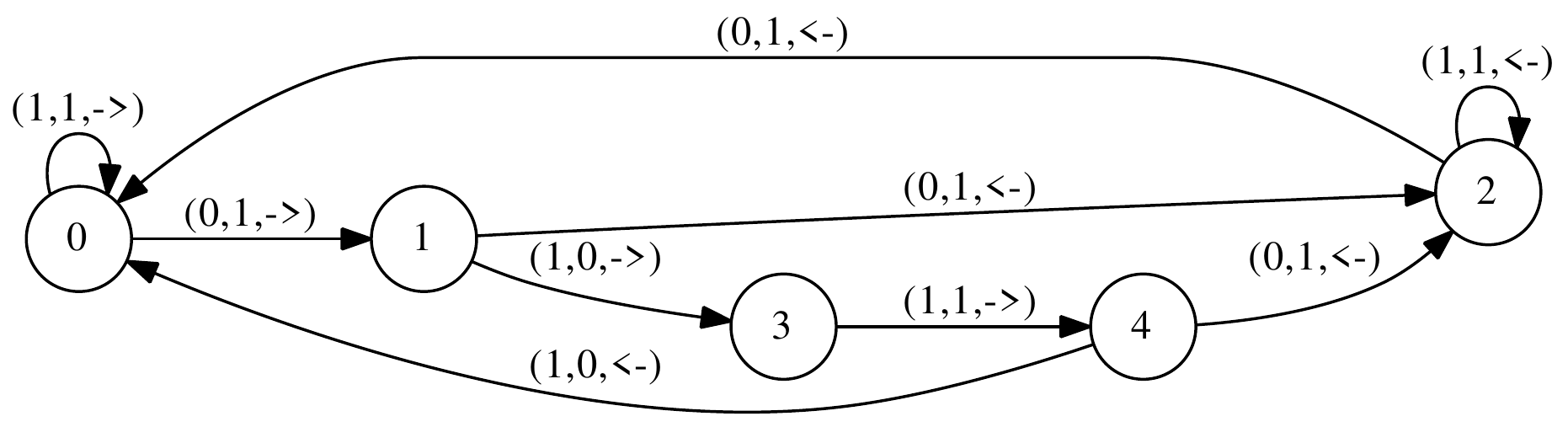}} 
\subfigure[4095, 11.809.985, (9, 0, 11, 1, 5, 2, 15, 3, 20, 4, 3, 5, 15, 7, 29, 8, 15, 9, 1)]{\label{recomb4095_2}\includegraphics[scale=0.33]{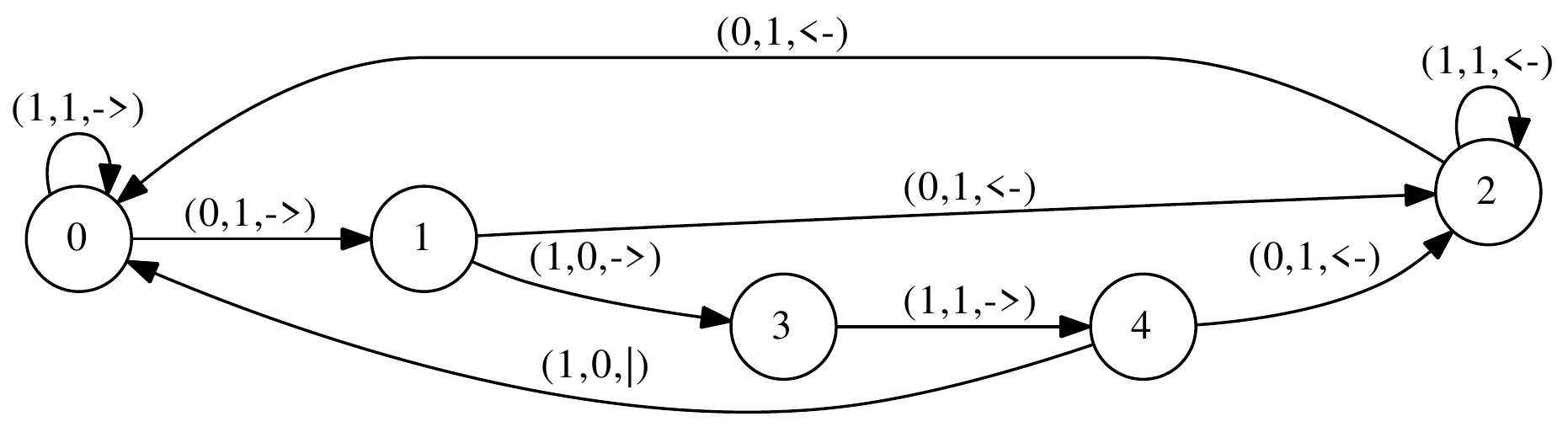}} \\
\subfigure[4095, 11.810.006, (9, 0, 11, 1, 5, 2, 15, 3, 20, 4, 3, 5, 15, 7, 29, 8, 24, 9, 1)]{\label{recomb4095_3}\includegraphics[scale=0.38]{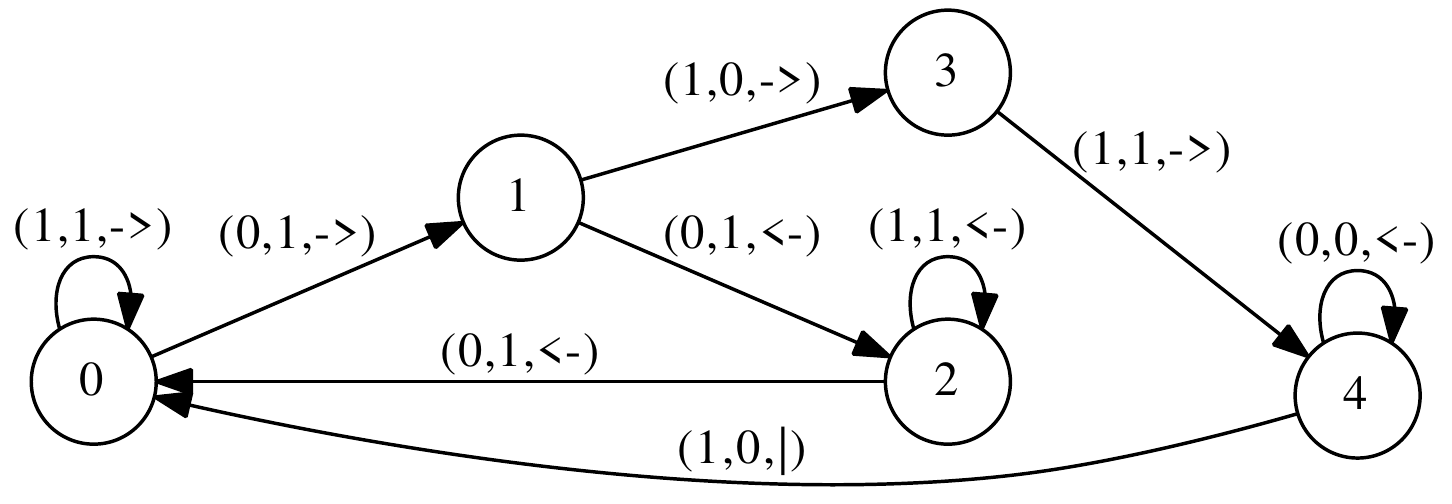}} 
\subfigure[4095, 11.821.189, (9, 0, 11, 1, 5, 2, 18, 3, 20, 4, 3, 5, 15, 7, 29, 8, 15, 9, 0))]{\label{recomb4095_4}\includegraphics[scale=0.35]{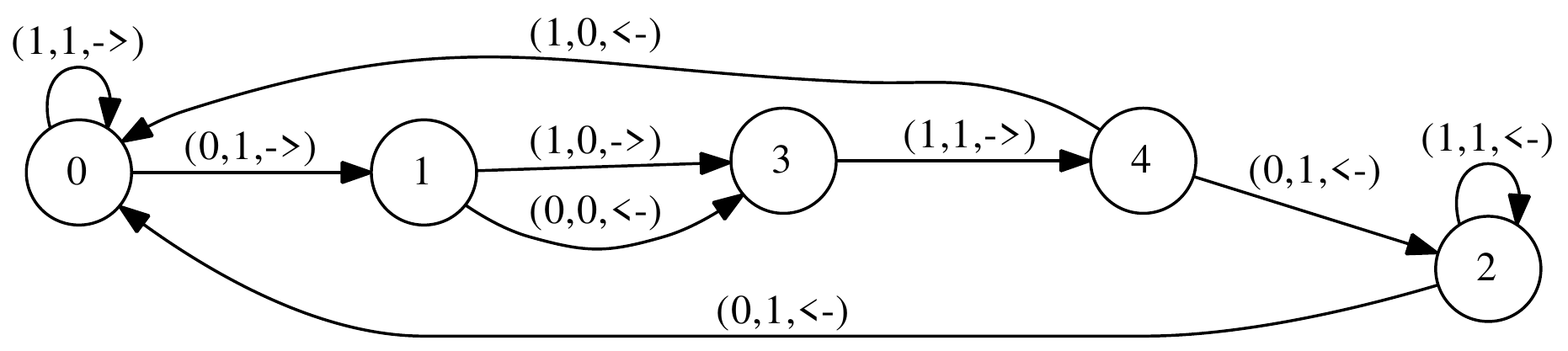}} \\
\subfigure[4095, 11.816.099, (9, 0, 11, 1, 5, 2, 18, 3, 20, 4, 3, 5, 15, 7, 29, 8, 15, 9, 1 ))]{\label{recomb4095_5}\includegraphics[scale=0.33]{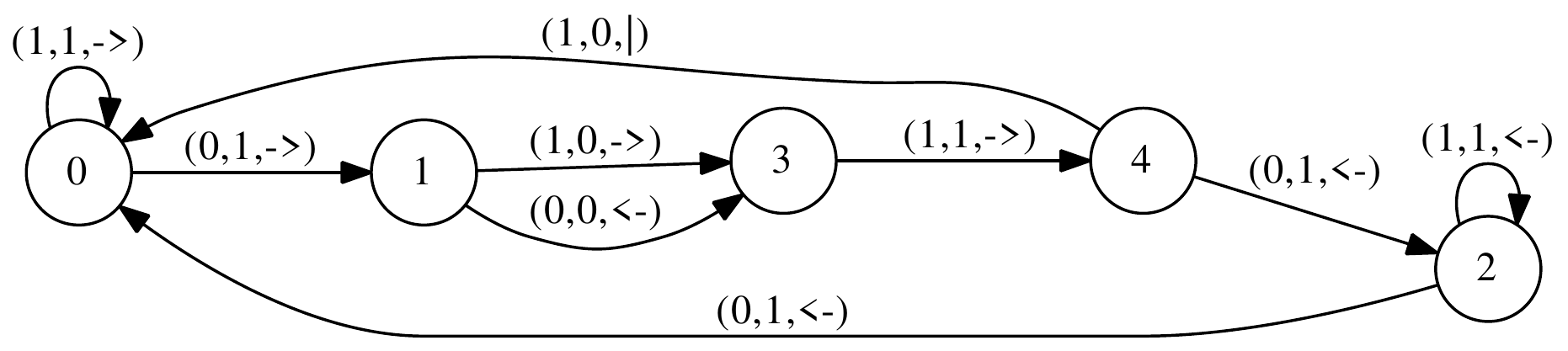}} 
\subfigure[4095, 11.811.009, (9, 0, 11, 1, 5, 2, 18, 3, 20, 4, 3, 5, 15, 7, 29, 8, 15, 9, 11))]{\label{recomb4095_6}\includegraphics[scale=0.33]{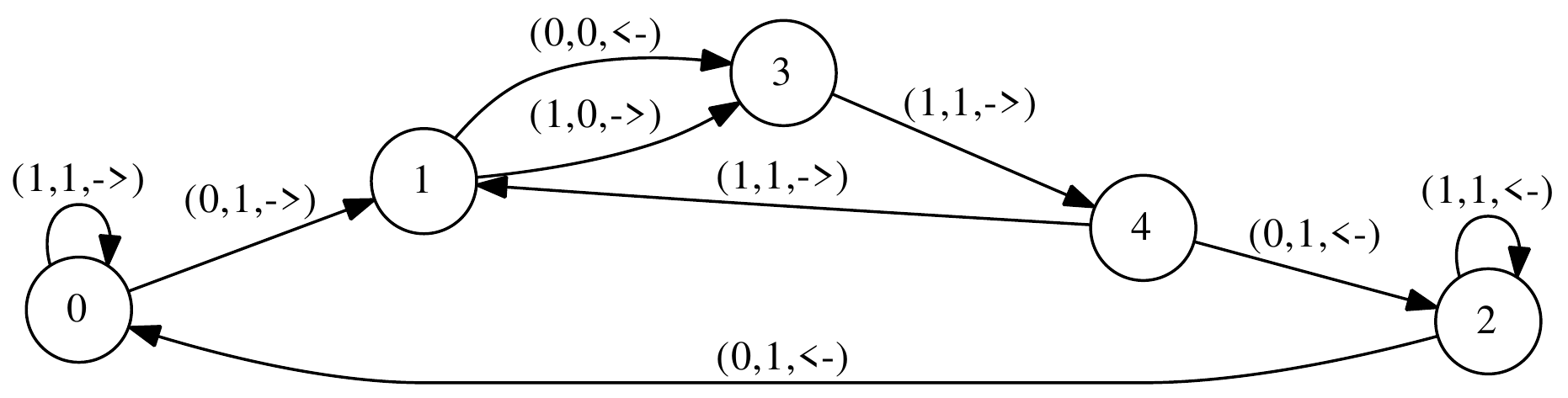}} 
\caption{Recombinations of $BB_5$ machines, $\mathcal{M}_{PP}(4095)$ machines, \# of 1's, \# of steps, name of the machine.}
\end{figure}

\clearpage
\section{Conclusion and further work}

In \citep[pp. 11]{michel-2009} Michel wrote that according to Marxen in range from $4096$ to $4098$ there are olny six machines which are listed in mentioned paper too. Hence, we were surprised that, using simply recombination methods, very similar, but different machines may be generated. For example, we have found a recombinated $BB_5$ machine which can make $70.740.809$ steps before halting. This machine was shown in Figure \ref{recomb4097_3} and their transition rules are shown in Figure \ref{winnercandidate}. The number $70.740.809$ may be the new winner candidate for Rad\'o $S$ fuction for $n=5$. We would get that $S(5)=70.740.810$ and $\sigma(5)=4098$ if we added the rule $(4,{\bf 0}) \rightarrow (H, {\bf 1},\rightarrow)$ to this recombinated machine.

\begin{figure}[htp]
\centering 
\begin{verbatim}
(0, 0)->(1, 1, 2)
(0, 1)->(2, 1, 0)
(1, 0)->(2, 1, 2)
(1, 1)->(0, 0, 1)
(2, 0)->(3, 1, 2)
(2, 1)->(4, 0, 0)
(3, 0)->(0, 1, 0)
(3, 1)->(3, 1, 0)
(4, 1)->(0, 0, 0)
\end{verbatim}
\caption{4097, 70.740.809, (9, 0, 11, 1, 15, 2, 17, 3, 1, 4, 23, 5, 24, 6, 3, 7, 21, 9, 0).}
\label{winnercandidate}
\end{figure}

\appendix

\section{Other recombinations}

Our recombinating and other related search programs are running. These programs and related data can be downloaded from \url{http://www.inf.unideb.hu/~nbatfai/bb}. We have already found machines from the following sets, 
$\mathcal{M}_{PP}(20)$,
$\mathcal{M}_{PP}(26)$,
$\mathcal{M}_{PP}(27)$,
$\mathcal{M}_{PP}(28)$, 
$\mathcal{M}_{PP}(30)$, 
$\mathcal{M}_{PP}(32)$, 
$\mathcal{M}_{PP}(39)$, 
$\mathcal{M}_{PP}(40)$,
$\mathcal{M}_{PP}(239)$. 
Finally we show some interesting $\mathcal{M}_{PP}(239)$ and $\mathcal{M}_{PP}(20)$ recombined machines.

\subsection{$\mathcal{M}_{PP}(239)$}

\begin{figure}[htp]
\centering                 
\subfigure[239, 41.082, (9, 0, 11, 1, 5, 2, 15, 3, 20, 4, 3, 5, 15, 7, 29, 8, 2, 9, 1)]{\label{recomb239}\includegraphics[scale=0.38]{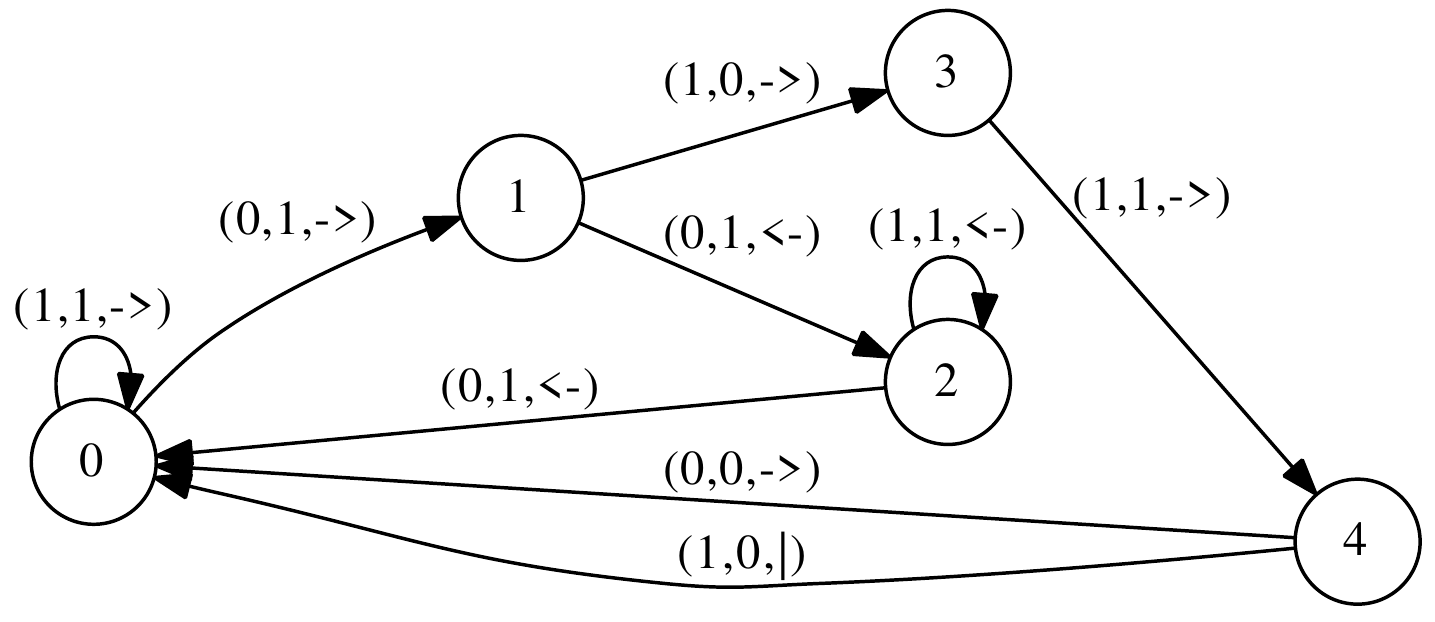}} %
\subfigure[239, 41.359, (9, 0, 11, 1, 5, 2, 15, 3, 20, 4, 3, 5, 15, 7, 29, 8, 2, 9, 0))]{\label{recomb239_2}\includegraphics[scale=0.38]{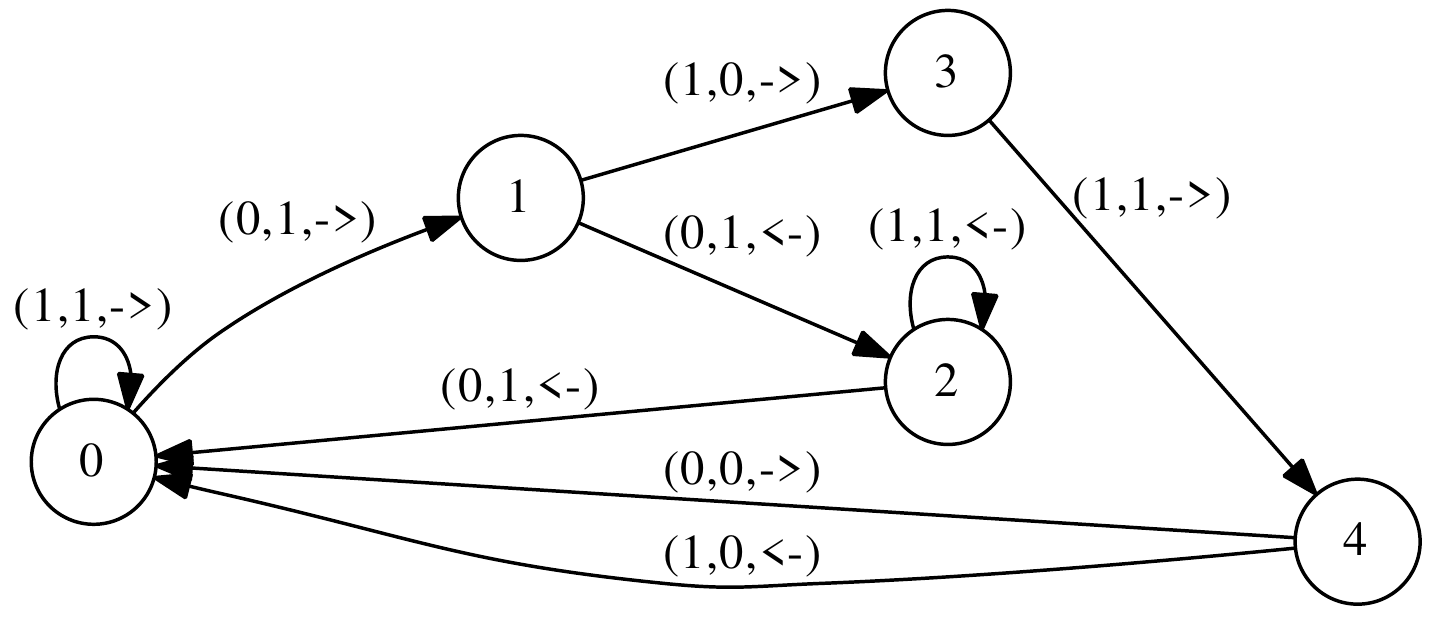}} 
\subfigure[239, 40.805, (9, 0, 11, 1, 5, 2, 15, 3, 20, 4, 3, 5, 15, 7, 29, 8, 2, 9, 11))]{\label{recomb239_3}\includegraphics[scale=0.38]{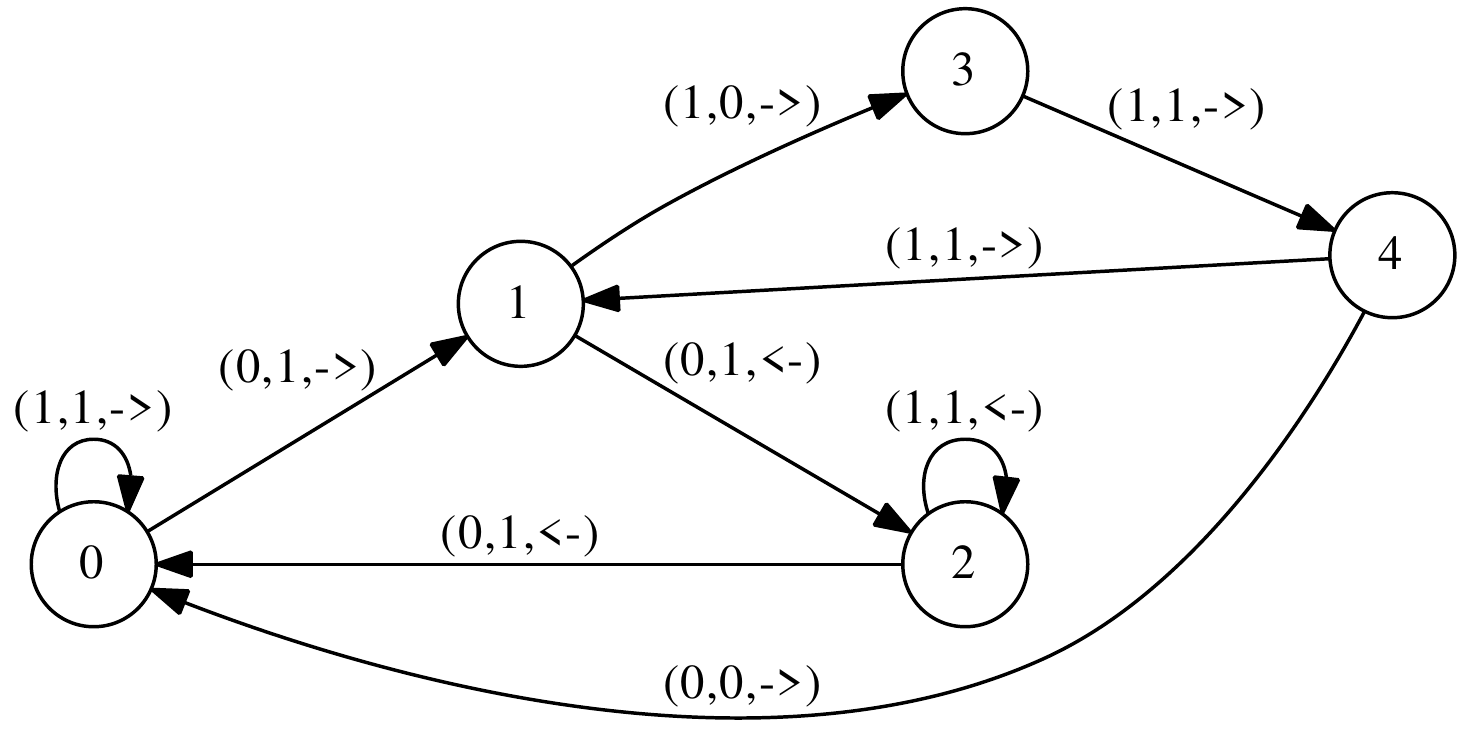}} 
\caption{Recombinations of $BB_5$ machines, two $\mathcal{M}_{PP}(239)$ machines, \# of 1's, \# of steps, name of the machine.}
\end{figure}

\clearpage
\subsection{$\mathcal{M}_{PP}(20)$}

\begin{figure}[htp]
\centering
\subfigure[20, 279, (9, 0, 11, 1, 5, 2, 15, 3, 23, 4, 3, 5, 15, 7, 29, 8, 15, 9, 8)]{\label{recomb20}\includegraphics[scale=0.33]{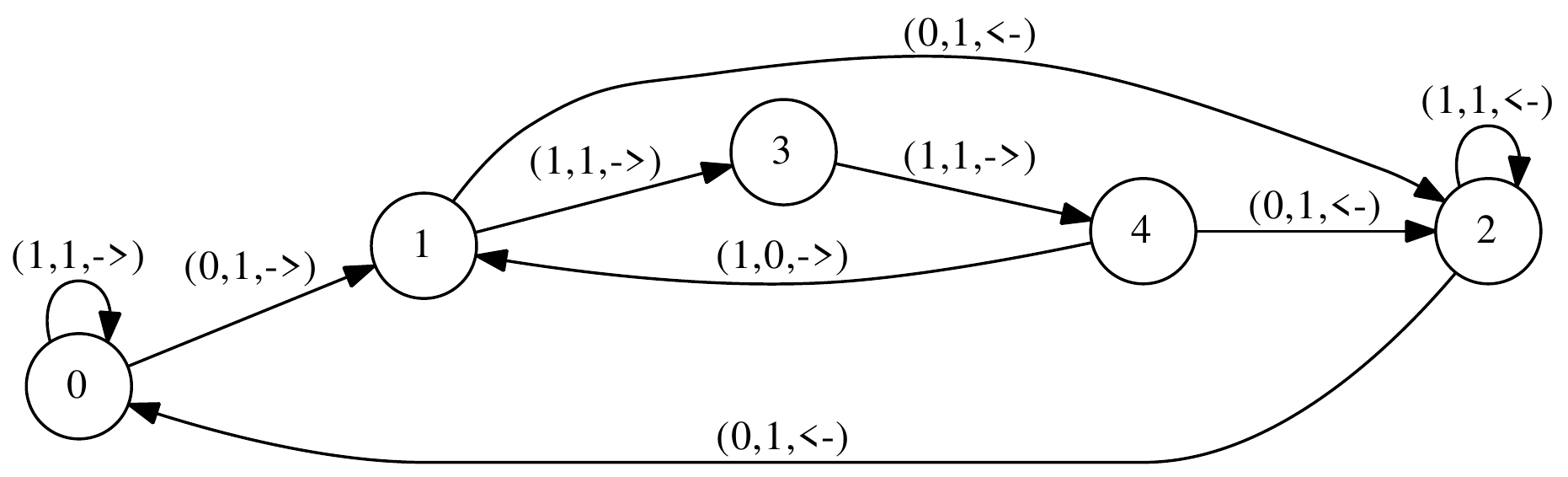}} %
\subfigure[20, 297, (9, 0, 11, 1, 5, 2, 15, 3, 23, 4, 3, 5, 15, 7, 26, 8, 0, 9, 11)]{\label{recomb20_2}\includegraphics[scale=0.33]{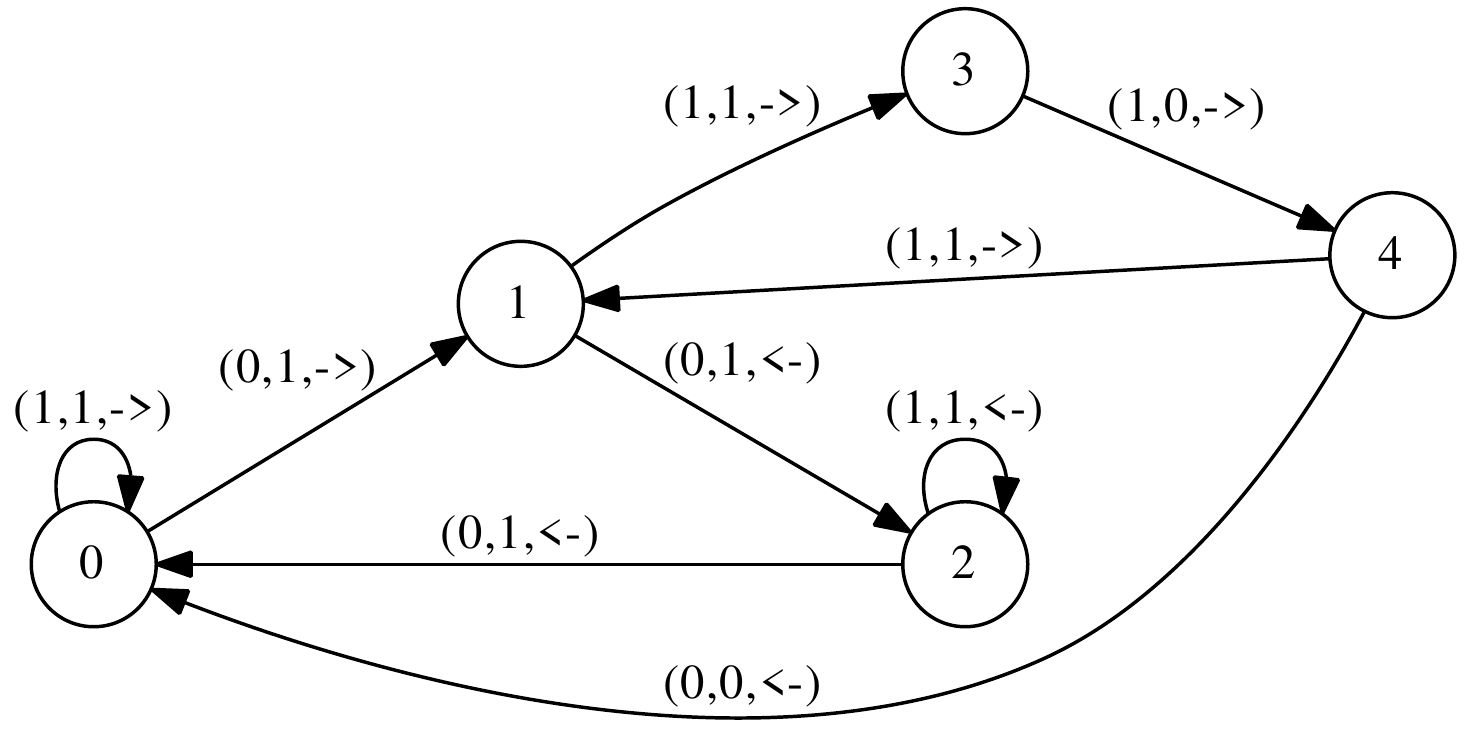}} 
\subfigure[20, 314, (9, 0, 11, 1, 5, 2, 15, 3, 23, 4, 3, 5, 15, 7, 26, 8, 0, 9, 1)]{\label{recomb20_3}\includegraphics[scale=0.33]{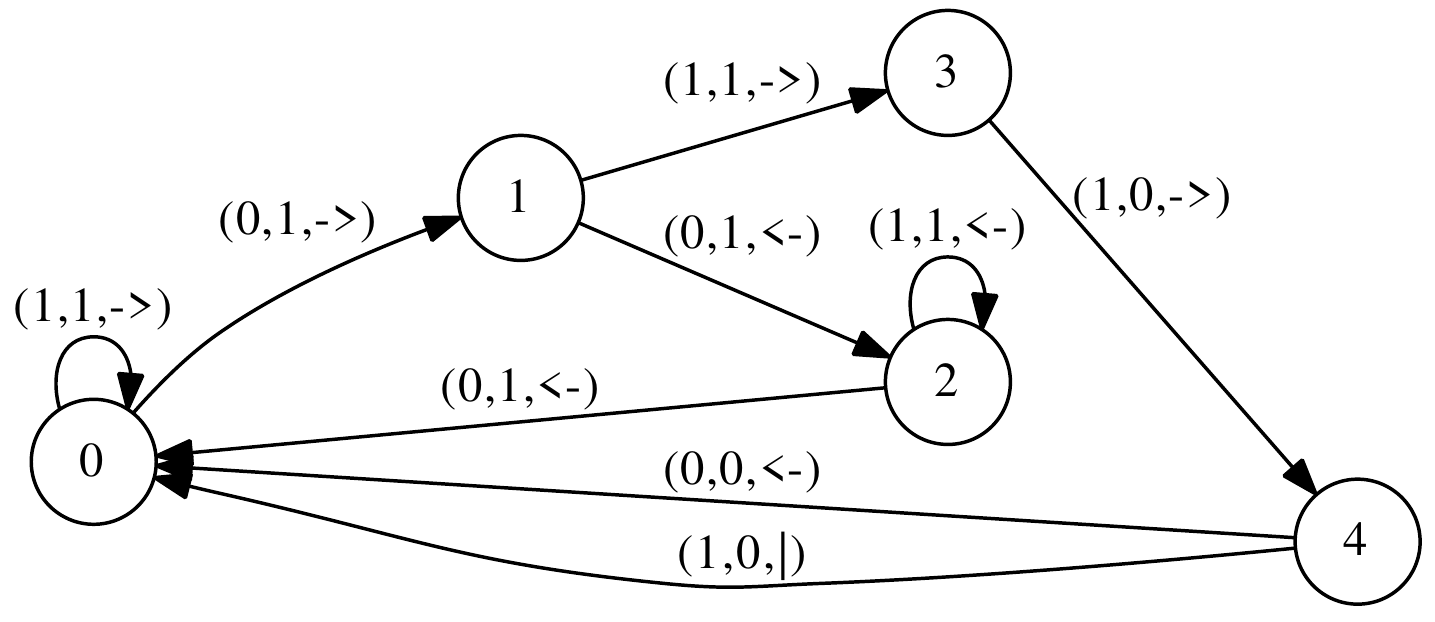}} 
\caption{Recombinations of $BB_5$ machines, a $\mathcal{M}_{PP}(20)$ machine, \# of 1's, \# of steps, name of the machine.}
\end{figure}

\clearpage
\section{Some C code snipets}

\subsection{The starting machines}
\label{codesnips}

The rows contained in this code snipet may be used to decrypt the recombinations of Turing machines. 
{\small
\begin{verbatim}
...
// Marxen-Buntrock, 4097,
{0, 11, 1, 15, 2, 17, 3, 11, 4, 23, 5, 24, 6, 3,  7, 21, 9, 0 }, // 0
// Marxen-Buntrock, 4096
{0, 11, 1, 18, 2, 15, 3, 23, 4, 3,  5, 15, 7, 29, 8, 5,  9, 8 }, // 1 
// Marxen-Buntrock, 4095
{0, 11, 1, 5,  2, 15, 3, 20, 4, 3,  5, 15, 7, 29, 8, 24, 9, 11 },// 2 
// Marxen-Buntrock, 4095
{0, 11, 1, 5,  2, 15, 3, 20, 4, 3,  5, 15, 7, 29, 8, 15, 9, 11 },// 3
// Marxen-Buntrock, 4097
{0, 11, 1, 5,  2, 15, 3, 9,  4, 5,  5, 21, 6, 5,  7, 27, 9, 12 },// 4
// Marxen-Buntrock, 4096
{0, 11, 1, 5,  2, 15, 3, 23, 4, 3,  5, 15, 7, 26, 8, 15, 9, 11 },// 5
// Uhing, 1471
{0, 11, 2, 15, 3, 17, 4, 26, 5, 18, 6, 15, 7, 6,  8, 23, 9, 5 }, // 6
// Uhing, 1915
{0, 11, 1, 15, 2, 0,  3, 18, 4, 3,  6, 9,  7, 29, 8, 20, 9, 8 }, // 7
// Schult, 501
{0, 11, 1, 12, 2, 17, 3, 23, 4, 3,  5, 8,  6, 26, 8, 15, 9, 5 }, // 8
// 160
{0, 9,  1, 12, 2, 15, 3, 21, 4, 29, 5, 1,  7, 24, 8, 2,  9, 27 },// 9
// 32
{0, 21, 1, 9,  2, 24, 3, 6,  4, 3,  5, 20, 6, 17, 7, 0,  9, 15 },// 10
// 26
{0, 9,  1, 11, 2, 17, 3, 21, 4, 19, 5, 29, 6, 5,  7, 6,  8, 8 }, // 11
// 21
{0, 9,  1, 11, 2, 15, 3, 20, 4, 21, 5, 27, 6, 4,  7, 2,  8, 12 },// 12
// 19
{0, 9,  1, 11, 2, 26, 3, 23, 4, 27, 5, 2,  7, 17, 8, 5,  9, 13 },// 13
...
\end{verbatim}
}

\clearpage

\bibliography{RecombinationsOf}

\end{document}